# Kerr nonlinearity effect on light transmission in one-dimensional photonic crystal

Daoud Mansour and Khaled Senouci*
University of Mostaganem Abdelhamid Ibn Badis,
Laboratoire de Structure, Elaboration et Application des
Matériaux Moléculaires (SEA2M), B.P. 227, Mostaganem, Algeria

**Abstract**

We investigate numerically the effect of Kerr nonlinearity on the transmission spectrum of a one dimensional δ-function photonic crystal. It is found that the photonic band gap (PBG) width either increases or decreases depending on both sign and strength of Kerr nonlinearity. We found that any amount of self-focusing nonlinearity ($\alpha > 0$) leads to an increase of the PBG width leading to light localization. However, for defocusing nonlinearity, we found a range of non-linearity strengths for which the photonic band gap width decreases when the nonlinearity strength increases and a critical non-linearity strength $|\alpha_c|$ above which the behavior is reversed. At this critical value the photonic crystal become transparent and the photonic band gap is suppressed. We have also studied the dependence on the angle of incidence and polarization in the transmission spectrum of our one-dimensional photonic crystal. We found that the minimum of the transmission increases with incident angle but seems to be polarization-independent. We also found that position of the photonic band gap (PBG) shifts to lower wavelengths when the angle of incidence increases for TE mode while it shifts to longer wavelengths for TM mode.

**Keywords:** Photonic crystals, Kerr effect, focusing nonlinearity, defocusing nonlinearity, photonic band gap, transmission spectrum, localization, delocalization, TE and TM polarization.

*Corresponding author email address: khaled.senouci@univ-mosta.dz

# I- Introduction

In the last three decades, photonic crystals (PCs) have attracted great interest and have been intensively studied both theoretically and experimentally due to their potential applications [1-6]. 1D Photonic crystals are periodic structures composed of alternating layers of material with high and low dielectric constants. The main features of PCs are the presence of forbidden frequency regions called photonic band gaps (PBGs) in its transmission spectrum. The electromagnetic waves with frequencies located in the PBG cannot propagate in the PCs [1,2].The formation of photonic band gap of conventional PCs results from multiple Bragg scattering of propagation waves which is strongly dependent on the incidence angle, lattice constant and polarization. The properties of light propagation can be controlled by varying the geometric and structural parameters of the PC. PCs have been applied in many aspects especially in optical devices such as optical diodes and switches, filters, wave guides, diode laser, photon polarization spectroscopy, and so on [7-11].

On the other hand, wave propagation in nonlinear media is a subject of intensive research both from the theoretical and experimental point of view [12-17]. In the linear regime the dielectric constant is independent of the field strength. However, the presence of optical nonlinearity in a system leads to a much richer and more complex response to radiation since the transmission coefficient is a function of the intensity of the incoming electromagnetic wave. Many interesting phenomena such as optical limiting and switching [18-21], optical bistability [22], self-trapping and stable localized modes [23, 24] are observed when nonlinear material response to light intensity is taken into account. One interesting and fascinating phenomenon of nonlinear photonic crystals is the existence of so-called gap solitons, discovered by Chen and Mills [25] in one-dimensional systems (1D). These gap solitons have a central frequency inside the linear photonic band gap. The existence of gap solitons was studied in GaAS/AlGaAs based one-dimensional nonlinear photonic crystal [26]. Nonlinear effects like Kerr effect, solitons etc.., play a more important role in advanced optoelectronics and photonics.

The optical Kerr effect related to the change in refractive index of the medium which is directly induced by the electric field of incident light. The importance of non-linearity of the PC has been shown when designing several non-linear devices which operate on the basis of the optical Kerr effect such as optical diodes [7, 27, 28], switches and limiters [29, 30].

Most works on 1D nonlinear PCs targeted to the Kerr nonlinearity effect on defect mode properties [31-33] whereas the treatment of its effect on the photonic band gap width has not been reported. In a recent paper [34], we studied the effect of a very weak defocusing Kerr nonlinearity on the transmission spectrum in 1D perfect photonic crystal. A defect mode-like peak, having a similar origin as the well known gap soliton, within the photonic band gap with a total transmission was observed. The purpose of this paper is to investigate numerically the effect of both strong self-focusing and self-defocusing Kerr nonlinearity on the transmission properties and particularly on the photonic band gap of one-dimensional -function photonic crystals (PCs). We will show the importance of the non-linearity strength on the behavior of the photonic band gap width. The influence of angle of incidence and polarization on the transmission spectrum has also been investigated.

# II- Model description

To investigate the effect of nonlinearity on the wave propagation in one-dimensional photonic crystal, we consider a one-dimensional Kronig-Penney model with N -function layers distributed periodically at *x=na, n=0,1…,N-1*. Here, a is the lattice constant. The schematic structure is shown in Figure 1. Such a periodic medium represents the simplest model for a one-dimensional PC.

The electromagnetic waves outside the nonlinear structure are described by [35]:

$$E(x) = \begin{cases} E_i e^{-ikx} + E_r e^{ikx}, & x \geq Na \\ E_t e^{-ikx}, & x \leq 0 \end{cases} \quad \text{………………………………(1)}$$

Here we considered an incident plane wave $E_i e^{-ikx}$ with wave number $k$, from the right which gives rise to a reflected wave, $E_r e^{ikx}$, as well as a transmitted wave $E_t e^{-ikx}$ where the wave vector *k=w/c, w* is the optical frequency, and *c* is the vacuum speed of the light.

Throughout this paper, only normal incidence will be considered. Inside the structure, the electric field for the transverse-electric (TE) mode satisfies the time-independent wave equation [15, 36]:

$$\frac{d^2 E(x)}{dx^2} + \beta \sum_{n=1}^{N} [1 + \alpha |E(x)|^2] E(x) \delta(x - Na) \text{…………………………(2)}$$

Here *E(x)* is the electric field at x-axis. $\beta = \frac{n^2 w^2}{c^2}$ where $n = \sqrt{\varepsilon}$ is the refractive index and $\varepsilon$ being the electric permittivity. The parameter $\alpha$ is the corresponding nonlinear Kerr coefficient. For simplicity the lattice spacing is taken to be unity throughout this work (*a=1*). Equation (2) is formally equivalent to the Kronig-Penney model of electrons [15, 41-43]. From the computational point of view it is more useful to consider the discrete version of this equation, which is called the generalized Poincaré map and can be derived without any approximation from equation (2). It reads [39]:

$$E_{n+1} = [2 \cos k - \beta(1 + \alpha |E(x)|^2) k. \sin k] E_n - E_{n-1} \text{…………………………(3)}$$

Where $E_n$ is the value of the electric field amplitude in polarization TE at site *n*. This representation relates the values of the electric field amplitudes at three successive discrete locations along the x-axis. The solution of the above equation is carried out iteratively by taking for our initial conditions the following values at sites *0* and *1*: $E_0=1$, and $E_1=e^{-ik}$. We consider here an electromagnetic wave having a wave vector *k* incident at site N from the right. The transmission coefficient *T* can then be expressed as [13, 40]:

$$T = \frac{4 \sin^2(k)}{|e^{-ik} E_n - E_{n-1}|^2} \quad \text{…………………………………………(4)}$$

Thus *T* depends only on the values of the electric field amplitude at the end sites, $E_n$ and $E_{n-1}$, which are evaluated from the iterative equation (3).

## III- Results and discussion

We perform numerical calculations to explore the Kerr nonlinearity effects on light propagation in a one-dimensional -function photonic crystal. Firstly, we discuss only the TE polarization of propagating electromagnetic waves at normal incidence. Equation (2), shows that with different nonlinear coefficients or different electric field intensity, the permittivity will have much different values, which may significantly influence the transmission property of the electromagnetic wave. In a defected photonic crystal, it was found that positive Kerr nonlinearity always shifts the defect modes toward longer wavelength, while the negative nonlinearity to shorter wavelength [32, 33]. The nonlinearity effect on the photonic band gap width of a perfect photonic crystal effect has not been studied.

First, let us examine how the photonic band gaps in one-dimensional photonic crystal are affected by Kerr nonlinearity. Both focusing and defocusing Kerr nonlinearity will be considered. In Figure 2, we show the effect of self-focusing ($\alpha>0$) and self-defocusing non-linearity ($\alpha<0$) on the transmission spectrum of a photonic crystal structure. We choose the number of the structure periods $N$ to be $6$. The refractive index of these layers is assumed to be $n=3.5$, which is the value for silicon. Figure 2a shows that for self-focusing non-linearity ($\alpha>0$), the PBG get shifted to higher wavelengths. As we increase Kerr nonlinearity strength, the PBG width increases. The PBG shift can be explained by the fact that a positive Kerr nonlinearity gives an increase for the refractive index and hence this change make a shift of the PBG position towards the higher wavelengths. Therefore, self-focusing nonlinearity tends to enlarge the photonic band gaps and thus localizes the propagating wave.

The situation is different if we consider a self-defocusing nonlinearity. The transmission spectra for a self-defocusing nonlinearity ($\alpha<0$) are displayed in Figure 2b. We found a critical value of Kerr nonlinearity $\alpha_c = -1$, for which the photonic crystal become transparent and a total transmission (zero gap) is observed. The suppression of the PBG for the critical value $\alpha_c$ can be explained by the fact that the parameter $\beta^* = \beta(1 + \alpha|E(x)|^2)$ in equation (2) tends to vanish for $\alpha_c = -1$. Two distinct behaviors for $|\alpha|<|\alpha_c|$ and $|\alpha|>|\alpha_c|$ are observed in Figure 2b. This figure shows that in the weak nonlinearity regime $|\alpha|<|\alpha_c|$, the PBG get shifted to lower wavelengths and its width decreases and get more and more suppressed as we increase the nonlinearity strength $|\alpha|$ (in magnitude). This figure shows also that for small values of self-defocusing nonlinearity, increasing field intensity $|E|$ reduces the effective parameter $\beta^*$ and the photonic band gap becomes narrower. Thus, the PBG width decreases with increasing values of Kerr nonlinearity provided that $|\alpha|<|\alpha_c|$. For larger values of the nonlinearity, $|\alpha|>|\alpha_c|$ (strong nonlinearity regime), the effect is reversed; that is the PBG width increases with nonlinearity strength $|\alpha|$.

To have a clear view of the effect of a defocusing Kerr nonlinearity, we consider a simple double photonic crystal structure. In figure 3, we presented the transmission spectrum of a double structure ($N=2$) for small values of both defocusing and focusing NL strength in comparison with the linear case ($\alpha = 0$). For each value of defocusing nonlinearity strength $|\alpha|$, we calculated the minimum of transmission $T_{min}$ (corresponding to the valley of the transmission, indicated by an arrow in figure 3). The results for weak nonlinearity regime $|\alpha|<|\alpha_c|$ are displayed in Figure 4. The self-defocusing nonlinearity delocalizes the electromagnetic wave since the minimum of the transmission increases with the amount of nonlinearity $|\alpha|$, while the self-focusing nonlinearity localizes the propagating mode (decrease of the minimum of the transmission). For the strong nonlinearity regime, Figure 5 shows that a self-defocusing nonlinearity can also localize the light above the critical value $|\alpha_c|=1$. In summary, the self-defocusing nonlinearity seems to decrease the PBG width and to shift its position to short wavelengths provided its strength satisfies $|\alpha|<|\alpha_c|$. When $|\alpha|>|\alpha_c|$ and for any amount of self-focusing nonlinearity($\alpha >0$), the PBG width increases and its position shifts to long wavelengths.

It is well known that the incident angle of light will influence the interference process within a PC. In Figure 6, we calculated the transmission spectra of the double structure at various values of angle of incidence ($\theta=0°,10°,20$ and $30°$) for a fixed amount of defocusing nonlinearity $\alpha = -0.1$. It is clear that the position of the photonic bandgap (represented by the valley) gradually shifts in the direction of short wavelength with the increase of the incident angle. It is also observed that the minimum of the transmission is affected by the variation of the incident angle. The transmission is enhanced with increasing incident angle. The same behavior of transmission dependence of the incidence angle was found in disordered photonic crystals [40]. It is also shown from this Figure that the minimum of the transmission increases with incident angle for a given nonlinearity strength. To examine the effect of polarization, we show in Figure 7 the minimum of the transmission $T_{min}$ and its corresponding wavelength $\lambda_0$ as a function of the incident angle for both TE and TM polarizations. For nonzero angle, the two polarization modes, TE and TM, possess different behaviors of $\lambda_0$. For TE mode, it decreases with the angle of incidence (PBG shifts to lower wavelengths) while for TM mode it increases (PBG shifts to higher wavelengths). In addition, the intensity of the minimum of transmission increases with increasing the angle of incidence. However, the variation of $T_{min}$ with the incidence angle seems to be polarization-independent (see inset of Figure 7).

## IV- Conclusion

In conclusion, we investigated in this paper the nonlinear response of wave propagation in a one-dimensional photonic structure. We have firstly studied the effect of self-focusing and self-defocusing Kerr nonlinearity on the transmission properties of one-dimensional (1D) photonic crystals (PCs) for TE polarization at normal incidence. We found that the photonic bandgap (PBG) is sensitive to the sign and strength of the Kerr nonlinear coefficient.

We have highlighted the role of nonlinear intensity on the behavior of the transmission. For self-defocusing nonlinearity, we found a range of non-linearity strengths ($|\alpha|<|\alpha_c|$) for which the photonic band gap width decreases when the nonlinearity strength increases and a critical non-linearity strength $|\alpha_c|$ above which the behavior is reversed. At this critical value the photonic crystal become transparent and the photonic band gap is suppressed. However any amount of self-focusing nonlinearity ($\alpha >0$) leads to an increase of the PBG width leading to photon localization.

Finally, we studied the transmission properties of our PCs under different incident angles for both transverse electric (TE) and transverse magnetic (TM) polarizations. We found that PBG is sensitive to incident angle and polarization of the incident light. When the angle of incidence increases, the PBG and the minimum of transmission were shifted toward the short wavelength regions for the TE mode while they were shifted toward the long wavelength regions for the TM mode. A significant increase in the intensity of the minimum of transmission is observed and seems to be polarization-independent.

We have studied the response of a nonlinear -function layer and it is interesting to extend this study to a nonlinear finite width layer. It is also interesting to study the competition between the nonlinear Kerr effect and the linear electro optic effect induced by the application of external electric field on our structure. This investigation will be the subject of a forthcoming paper.

**Acknowledgments**

One of the authors (K. S.) would like to acknowledge the International Centre for Theoretical Physics (ICTP, Trieste, Italy) for its hospitality where a part of this work was done. The author also gratefully acknowledges support from "La Direction Générale de la Recherche Scientifique et du Développement Technologique" (DGRSDT). This work has been supported by "Projet de Recherche Formation Universitaire" PRFU (Grant No B00L02UN270120180001 approved in 2018).

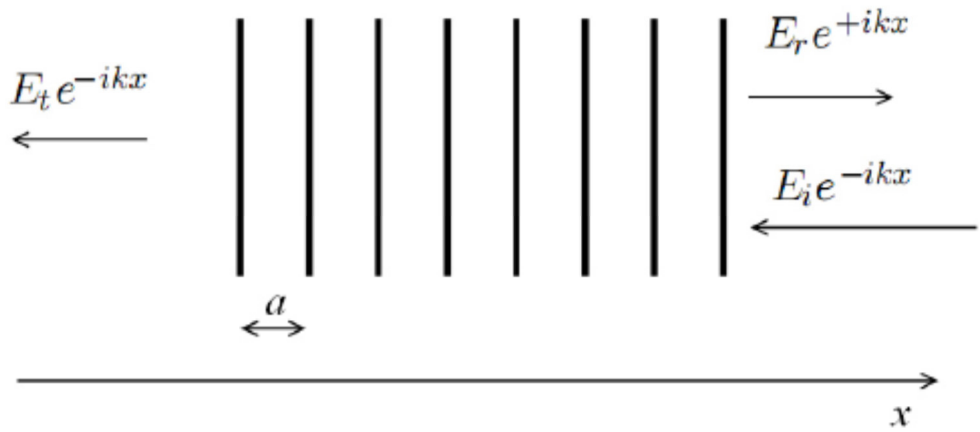

**Figure 1:** One-dimensional system of N -function layers separated by lattice spacing a (air).

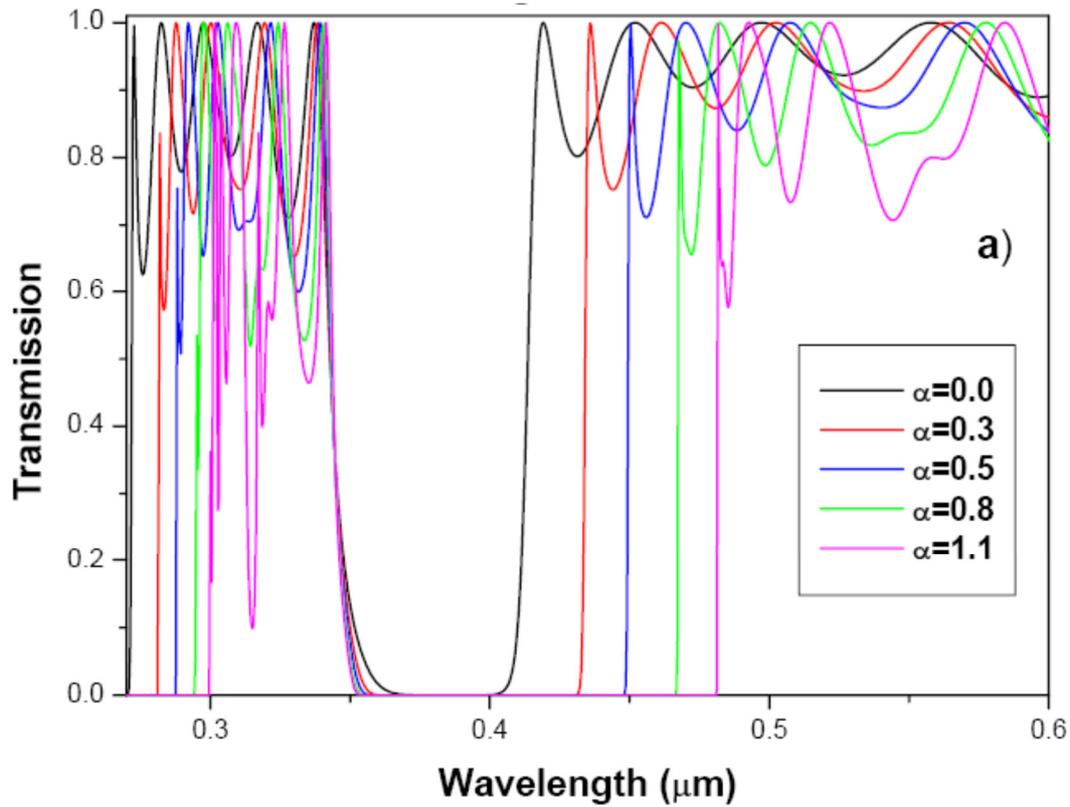

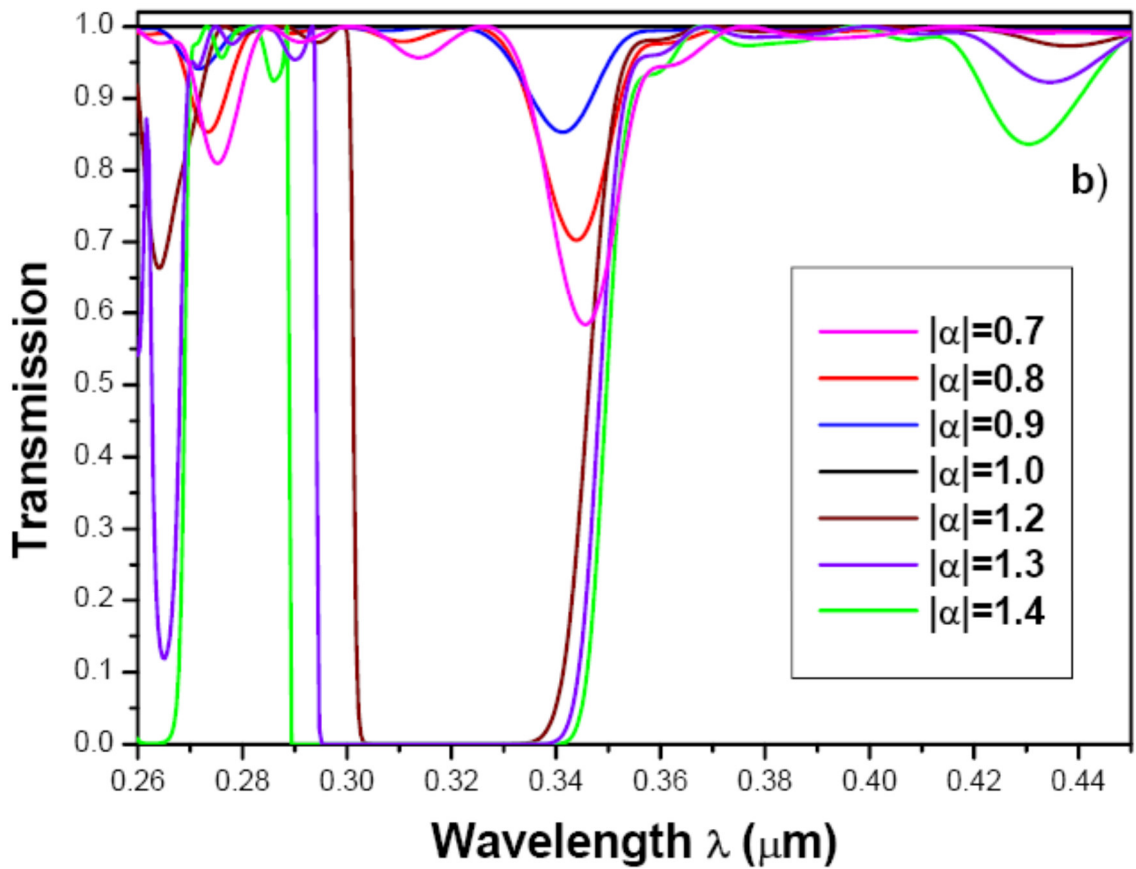

**Figure 2:** (Color online) Transmission spectra as a function of the wavelength for a structure of 6 layers and a refractive index n=3.5. a) Effect of self-focusing nonlinearity ($\alpha$ >0). b) Effect of self-defocusing nonlinearity ($\alpha$ <0).

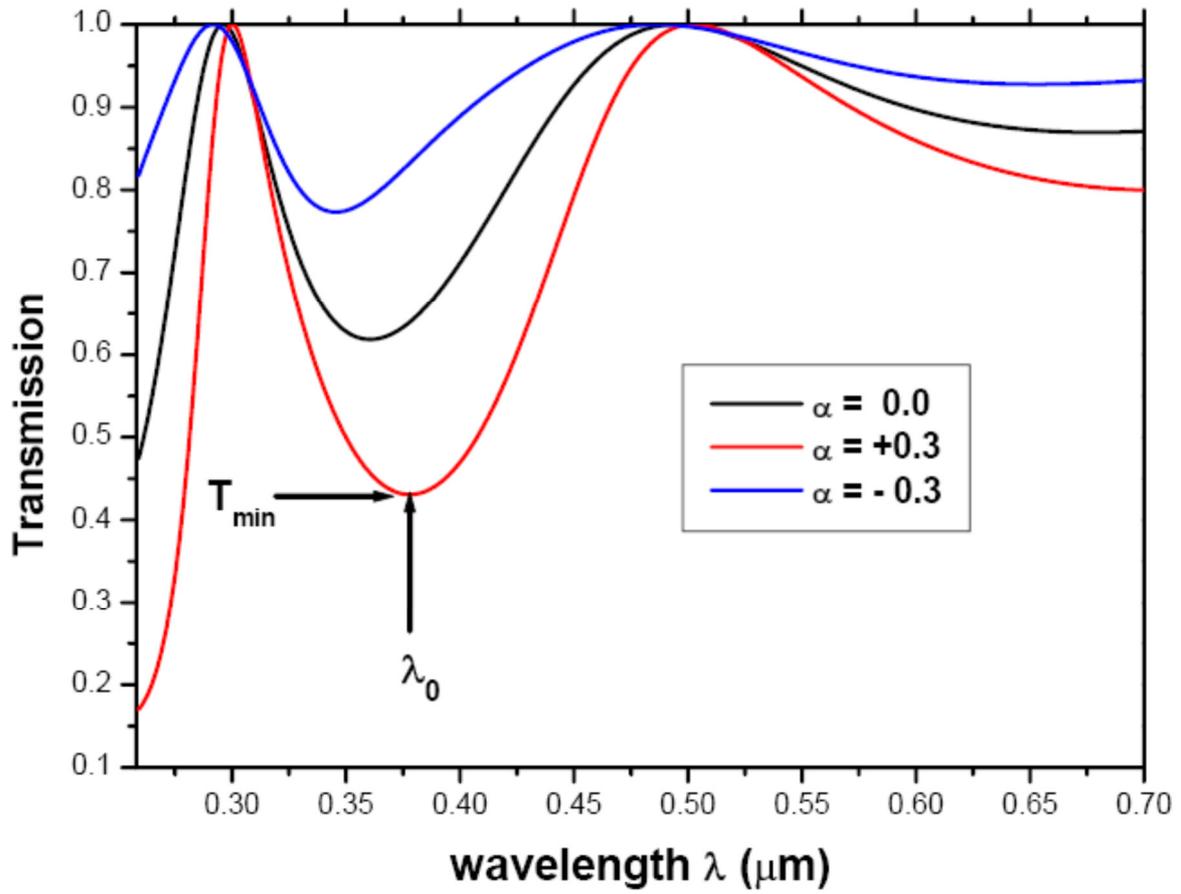

**Figure 3:** (Color online) Transmission spectra as a function of the wavelength for a double structure for TE wave at normal incidence for $\alpha = 0, -0.1$ and $+0.1$.

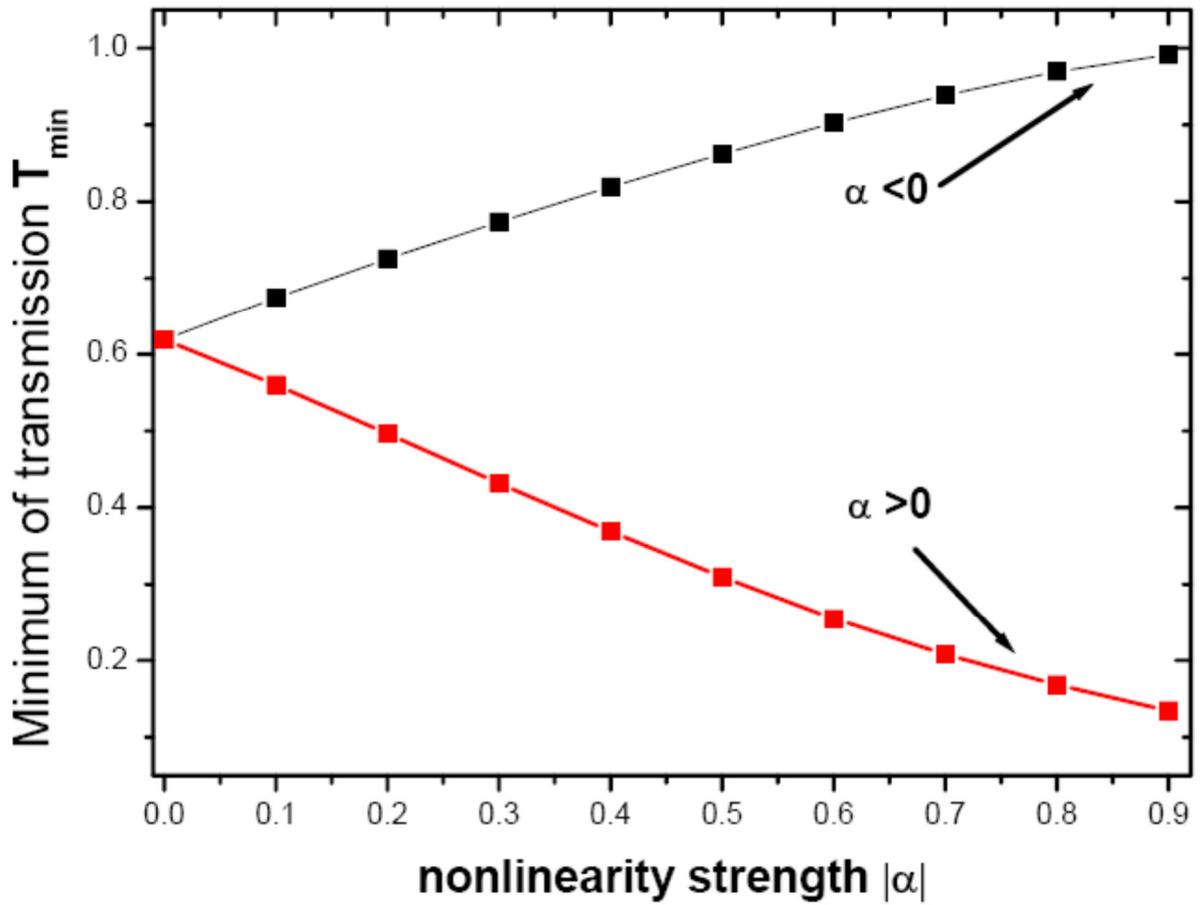

**Figure 4:** (Color online) Minimum of the transmission as a function of nonlinearity strength for a double structure for $|\alpha|<|\alpha_c|$ (weak nonlinearity regime) for TE wave at normal incidence. The solid lines are guides for the eyes.

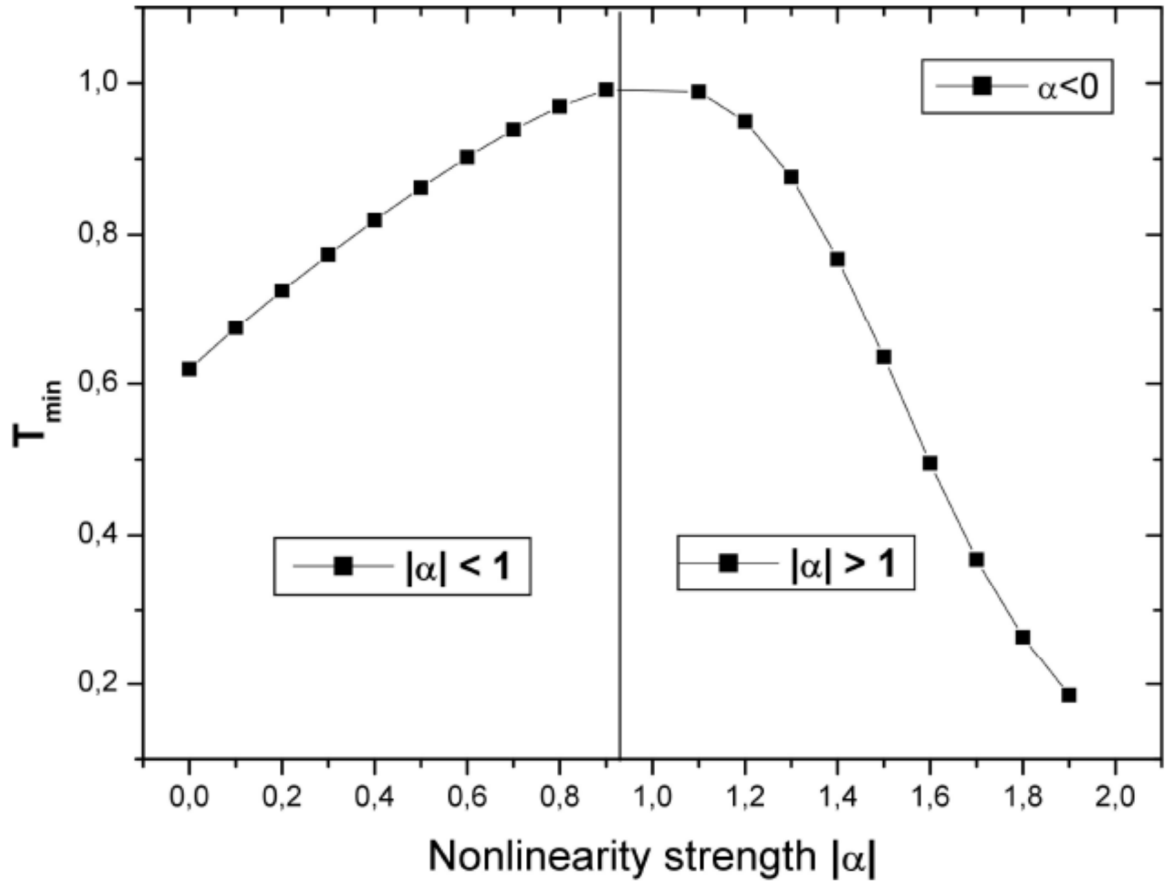

**Figure 5:** Minimum of the transmission as a function of nonlinearity strength for a double structure for a self-defocusing nonlinearity $\alpha <0$, weak and strong nonlinearity regimes for TE wave at normal incidence $\theta =0°$. The solid lines are guides for the eyes.

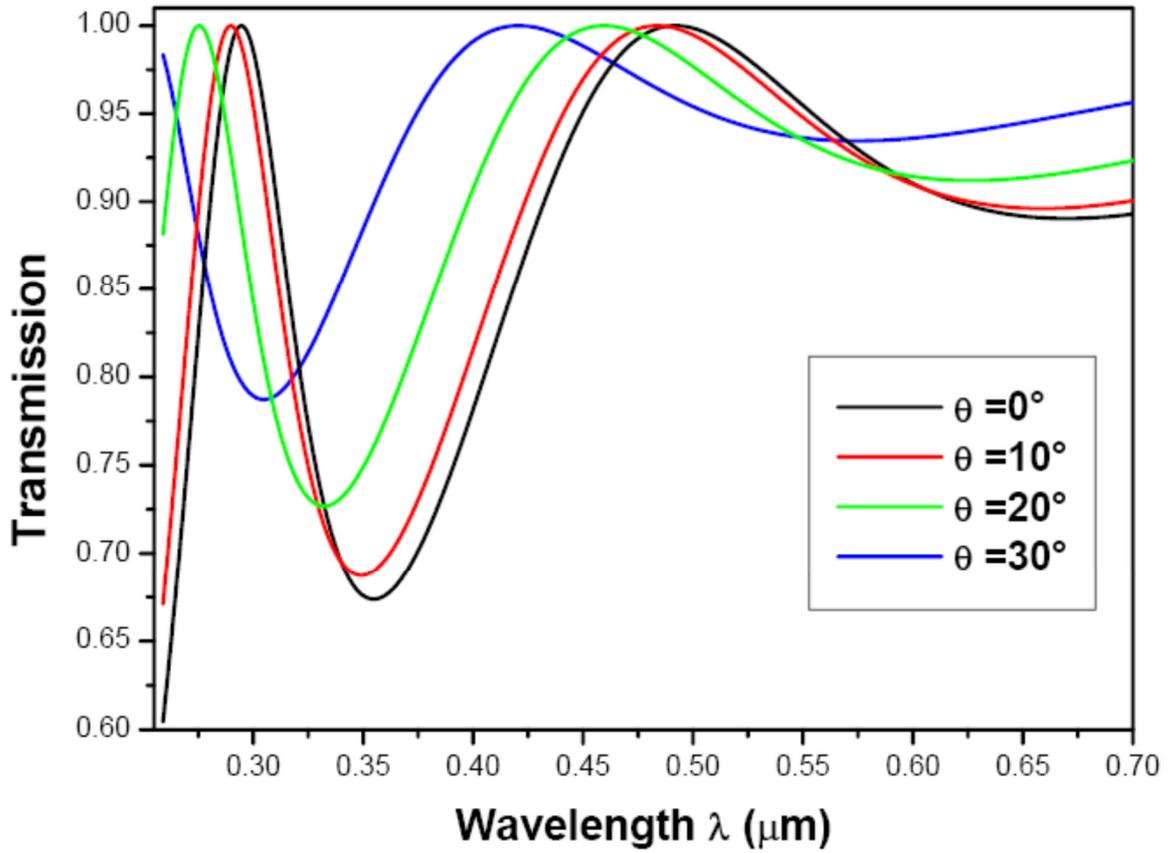

**Figure 6:** (Color online) Transmission spectra as a function of the wavelength for a double structure at various values of angle of incidence (θ=0°, 10°, 20°and 30°) for TE wave for a defocusing nonlinearity $\alpha$ =-0.1.

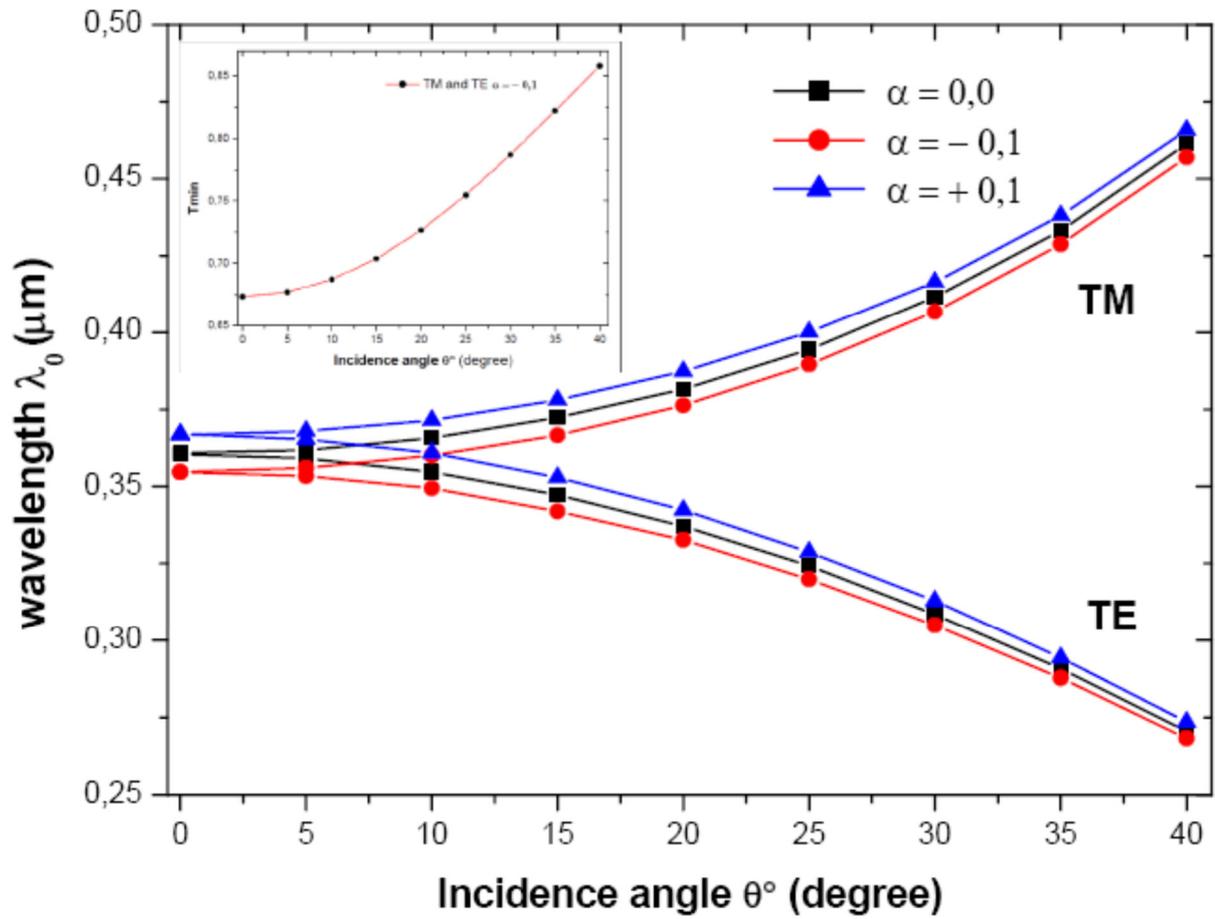

**Figure 7:** (Color online) The minimum transmission wavelength $\lambda_0$ as a function of the angle of incidence for both TE and TM mode and for *α =0,-0.1 and +0.1*. Inset: Minimum of the transmission as a function of angle of incidence for *α =-0.1*. The solid lines are guides for the eyes.